\documentclass[preprint,11pt, a4paper]{elsarticle}
\usepackage{amssymb}
\usepackage[margin=2.6cm]{geometry}

\usepackage{lineno}
\usepackage{tikz}
\newcommand*\circled[1]{\tikz[baseline=(char.base)]{
            \node[shape=circle,draw,inner sep=2pt] (char) {#1};}}
     
\usepackage{algorithm}
\usepackage[noend]{algpseudocode}
\usepackage{graphicx}
\usepackage{grffile}
\usepackage{longtable}
\usepackage{wrapfig}
\usepackage{rotating}
\usepackage{amsfonts}
\usepackage[T1]{fontenc}
\usepackage{courier}
\usepackage{amsmath} 
\usepackage[normalem]{ulem}
\usepackage[english]{babel}
\usepackage[utf8]{inputenc}
\usepackage{fancyhdr}
\usepackage{epstopdf}
\usepackage{booktabs}
\usepackage{hyperref}
\usepackage{multirow}

\journal{SoftwareX}

\begin{document}

\begin{frontmatter}



\title{ \huge \textbf{Founsure 1.0}: An Erasure Code Library with Efficient Repair and Update Features}


\author{ \Large \c{S}uayb \c{S}. Arslan}

\address{ \large MEF University, Maslak, Istanbul, Turkey}

\begin{abstract}
Founsure is an open-source software library that implements a multi-dimensional graph-based erasure coding entirely based on fast exclusive OR (XOR) logic. Its implementation utilizes compiler optimizations and multi-threading to generate the right assembly code for the given multi-core  CPU architecture with vector processing capabilities. Founsure possesses important features that shall find various applications in modern data storage, communication, and networked computer systems, in which the data needs protection against device, hardware, and node failures. As data size reached unprecedented levels, these systems have become hungry for network bandwidth, computational resources, and average consumed power. To address that, the proposed library provides a three-dimensional design space that trades off the computational complexity, coding overhead, and data/node repair bandwidth to meet different requirements of modern distributed data storage and processing systems. Founsure library enables efficient encoding, decoding, repairs/rebuilds, and updates while all the required data storage and computations are distributed across the network nodes. 
\end{abstract}


\end{frontmatter}



\section{Motivation and significance}
\label{}

Erasure coding is a fault tolerance mechanism that provides data protection and high availability in distributed data storage and processing systems \cite{netapp}. Reed-Solomon (RS) codes are the conventional option for constructing erasure codes based on overhead--optimal design. In other words, they use the storage space as efficiently as information-theoretically possible \cite{RS}. As the modern data storage systems evolved to possess different requirements, the set of constraints on the design of erasure codes has dramatically changed. For instance, previous research work on erasure coding such as RS focused on optimizing the coding overhead i.e., minimization of storage space for a given target data reliability  \cite{RS, Cauchy}. Moreover, some of the most popular designs considered pure eXclusive Or (XOR) operations to provide durability, and efficient computation \cite{evenodd}. More recently, locally repairable codes have attracted attention due to their efficient utilization of network resources and eventually achieve better overall reliability \cite{XORing} at the expense of suboptimal coding overhead. Besides the proprietary implementations of advanced erasure coding algorithms, many open-source implementations with different mathematical constructions become available online \cite{partow, Plank}. Previous works provided overhead optimal and fast/efficient erasure coding library functions. However, they did not take into account the peculiarities of the harnessed network, scarce computational resources, and data/node regeneration in a distributed setting. 

Founsure utilizes binary operations on a three-dimensional bipartite graph to construct a multi-functional erasure code. The design space includes computational complexity, coding overhead, and repair bandwidth as different dimensions of optimization. If one prefers to have an overhead optimal design with a reasonable complexity performance, then it might be advisable to use the well known Jerasure 2.0 \cite{Plank} erasure coding library, which is now fully supported by the \textit{RedHat} \textit{Ceph} community \cite{ceph_link}. Unfortunately, this library and the likes (e.g. zfec \cite{zfec}) do not provide a sufficient code structure to address modern problems of distributed storage systems such as degraded reads, data repair/regeneration bandwidth, data security, etc. On the other hand, the main objective and purpose of developing the Founsure library have been to provide different operating points in the three-dimensional design space based on the requirements of the storage applications through a set of parameter configurations. Therefore, Founsure can be shown to demonstrate the huge potential for distinct storage applications using smart/guided configuration steps. For instance, Founsure is shown to be configured to a baseline deduplication engine within an archival scenario \citep{Suayb1}.

The encoding process of Founsure begins by defining a two-dimensional conventional bipartite graph which leads to a non-systematic low-density generator matrix (LDGM) code. A version of this class of codes, with appropriate input distributions, are generally known as fountain codes \cite{mckay, luby}. The degree distribution of Founsure's LDGM code is specially selected to meet a good trade-off operating point between computational complexity, coding overhead, and repair bandwidth. The current version (1.0) supports Robust Soliton Distribution (RSD) \cite{luby} (if one prefers good coding overhead design), as well as all possible finite max-degree distributions (if one prefers different operating points) including the one in \cite{amin} by default. However, we note that the referred distributions are optimized for the minimum coding overhead criterion only.

One of the building pillars of Founsure is genuine symbol check relationships. The terminology of check symbols is quite common in Low Density Parity Check (LDPC) coding community. What check symbols do is that they provide a mathematical relationship between a subset of data symbols to check a certain condition, usually a simple binary sum or equivalent Galois field operation. The check idea is also quite beneficial for local data repairs/rebuilds \cite{repairLDPC}. So on top of the two-dimensional bipartite graph of Founsure, the encoding engine generates check nodes for ``data-only'' (referred hereafter as check \#1), ``data \& coding'' (referred as check \#2) and ``coding-only'' (referred as check \#3) chunks/symbols. As can be imagined, these check nodes (mathematical relationships) can be added to the two-dimensional bipartite graph to give it a three-dimensional look. This new data representation shall be used to provide advanced decoding, repair, and update features of the library. Throughout the document, nodes typically contain multiple chunks and chunks typically contain multiple symbols.  

Founsure uses Belief Propagation (BP) \cite{Pearl} (a.k.a. message passing) algorithm to resolve or decode the user data, to repair the encoded data, or update the encoded data. Sticking to BP as a design criterion is to ensure a low-complexity decoding process and allow fast/efficient operation. Library also supports register-level parallelism through compiler optimizations as well as multi-threading using the open standard \textit{openMP} primitives with its encode, decode, repair and update functions. The multi-threading feature, once properly configured and used, allows parallel processing and ensures acceleration for shared-memory architectures.  By reducing the processing time, Founsure secures quick responses to the common read/write requests of any generic distributed storage system. 

With the current release, the original user data does not appear in the output files. Instead, all output files are a mathematical function of the user data due to the so-called non-systematic encoding. In other words, one cannot read off data from the encoder output without any further decoding. Therefore, using Founsure, one can think of the user data encrypted automatically after encoding operation. Note that we use pseudo-random number generators (based on linear congruential generator) and seed (integers of long type) to build edges of the underlying bipartite graph. So without the seed number (we can treat them as keys in an encryption context), there is no way to recover original user data simply because the underlying graph generation is contingent upon the seed availability. Therefore, the Founsure software package also provides a user-configurable lightweight built-in encryption feature. Unlike systematic codes in which the data is explicit at the encoder output, Founsure comes with the non-systematic format as mentioned before. This, however, provides data security in return in addition to data protection.  As long as decoding is fast using various novel techniques we discuss in this study, non-systematic codes should not be an overall performance bottleneck for the rest of the system. 

This paper shall briefly describe the details of the software architecture,  a set of functionalities provided with the library as well as some of the associated advanced features. The source code, a comprehensive user guide, few test results, and all related documentation are available also from \href{https://https://github.com/suaybarslan/founsure}{\textit{github}}  and the web link \url{http://www.suaybarslan.com/founsure.html}.



\section{Software description and architecture}
\label{}

\subsection{Software Functionalities}
\label{}

Founsure has the following three executable main components that implement four important functionalities.
\begin{itemize}
\item founsureEnc: Encoder engine that generates $s$ number of data chunks (to be stored in $s$ different failure domains) under a local \textit{Coding} directory and a metadata file that includes information about the file, coding parameters, and the seed information.
\item founsureDec: Decoder engine that requires a local \textit{Coding} directory with enough number of files, a valid file name, and an associated metadata file to run multiple Belief Propagation (BP) passes in order to decode the user data. 
\item founsureRep: Repair engine that also requires a \textit{Coding} directory with sufficient number of files and
\begin{itemize}
\item fixes/repairs one or more data chunks should they have been erased, corrupted, or flagged as unavailable. 
\item generates extra coding chunks should a code update has been requested.  The system update is triggered if data reliability is decreased/degraded overtime or increased due to equipment replacements. 
\end{itemize}
\end{itemize}

These functions are used to execute encoding, decoding, repair, and update operations. There are also utility functions of Founsure used to help system admins to make correct design choices on degree distributions, required reliability, desired complexity, and storage space efficiency. We also use utility functions to trigger update functionality as will be demonstrated later. One of the distinctive features of utility functions is that they do not directly process user data, instead, they help us configure right parameters for the main functions to modify and process the user data properly. The current version supports two utility functions as listed below. 

\begin{itemize}
\item {simDisk}: This function can be used to exhaust all possible combinations of disk failures for a given set of coding parameters. In other words, this function checks whether the provided coding parameters are sufficient to achieve a user-defined reliability goal. Therefore, running this function can help us design target-policy erasure codes by configuring degree distributions for achieving various system-level goals besides reliability. 
\item {genChecks}: This utility function is crucial for two different important functionalities: (1) fast/efficient repair/rebuild of data and (2) seamless on-the-fly update. For the repair process, it generates two types of checks: check \#2 and check \#3 and registers them into a <testfile>\_check.data file using a format described within this document. In case of an update, it modifies the metadata and <testfile>\_check.data files so that the coding chunks can be updated by running {founsureRep} function. 
\end{itemize} 

Next, we provide the details of Founsure encoding, decoding, repair and update operations, particularly the implementation details of {founsureEnc}, {founsureDec} and {founsureRep} functions. For more details on underlying theory and computational complexity, we refer the reader to the appropriate documents such as \cite{sarslan1}.

\subsection{Implementation details of Encoding/Decoding Operations}
\label{}

In graph theory terminology, nodes (sometimes referred to as equations) are represented by graph vertices and node relationships by edges of the graph. There are three types of nodes in a 3-D bipartite graph; \textit{data} nodes, \textit{coding} nodes and \textit{check} nodes. The coding nodes represent a set of linear combinations of data nodes generated through a predetermined mathematical function such as XOR logic operation. Check nodes represent all the local sets of data and coding nodes for which a certain mathematical relationship is satisfied such as even or odd \textit{parity}. A simple mathematical function used by Founsure is the region XOR operation that operates over multiple data blocks of the same size and generates a single block of information. We use $f$ flag to indicate the file name, $k$ to indicate the total number of data nodes/symbols where $b$ of these are the original user data nodes/symbols, $n$ to indicate the total number of coding nodes/symbols, and $t$ to indicate the number of bytes to store per node/symbol. 

\begin{figure}[t!]
\centering
\includegraphics[width=0.8\textwidth]{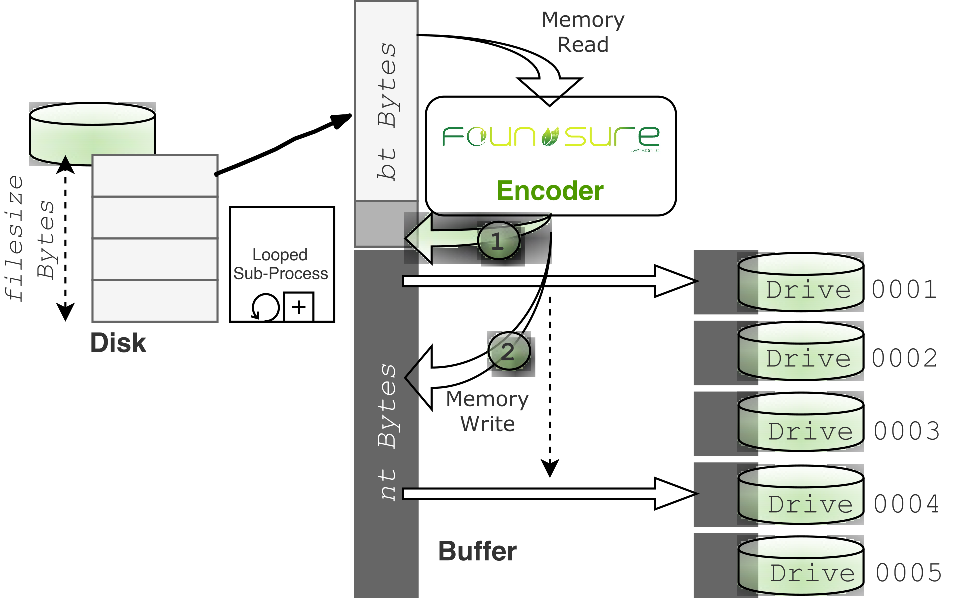}
\caption{The function {founsureEnc} allocates $(n+k)t$ bytes of buffer, generates coding chunks, and write them to a shaded area on memory before they are written to distinct drives. The file on disk is striped and processed in a looped subprocess as shown. The number of stripes is stored in the $readin$ variable and written to the metadata file.} \label{encoder_struc}
\end{figure}

In {founsureEnc} function, data file with \textit{filesize} bytes is partitioned into multiple $b \times t$ bytes and each partition is encoded independently as shown in Figure \ref{encoder_struc}. With the current version, \textit{partition coupling} is not supported between distinct partitions i.e., partitions are processed independently of each other. This technique is currently under investigation and might have interesting performance improvements to our design/implementation in analogy to spatially-coupled LDPC codes \cite{SCLDPC}. However, the coupling may have different effects for partial disk failures and may eventually lead to non-uniform decoding performances across partitions. 

If \textit{filesize} is not a multiple of $b \times t$ bytes, then we use zero padding to make \textit{filesize} a multiple. On the otherhand, {founsureEnc} also checks whether $s$ divides $n$ ($s | n$). If not, the least largest $n$ is selected automatically to satisfy  $s | n$. Such a requirement enables us to store exact same amount of information bytes across different failure domains.  This is particularly relevant to a balanced system design where the underlying storage devices are of the same type, quality and equally resilient against various types of failures.


Encoding proceeds as follows. First, a memory space (a buffer) worth $(k+n)t$ bytes is allocated, and the buffer content is initialized to zero. Next, check \# 1 equations are generated by an efficient array LDPC encoding \cite{arrayldpc}, \cite{ArrayLDPC1}. The choice of array LDPC as the precode is to enable efficient encoding operation and fast processing. As a result of this operation, an extra $k-b$ chunks are created to make up a total of $k$ chunks of data. The precoding process is shown as \circled{1} in Figure \ref{encoder_struc}. Later, a total of $n$ coding chunks are generated from the whole set of $k$ data chunks based on an LDGM base code with a configured "FiniteDist" degree and pseudo-random selection distributions. This process is shown as \circled{2} in Figure \ref{encoder_struc}. Finally, $n$ coding chunks are distributed (striped) equally across distinct output files for allocation on \textit{s} number of drives. We repeat this process for each data partition in a loop and append coding chunks at the end of the corresponding output files. For a given \textit{<filename>.ext} file, we use \textit{<filename>\_disk{0..0i}.ext} to refer to the $i$th output file. The number of zeros that appear in the name of output files is set by the "parameter.h" variable DISK\_INDX\_STRNG\_LEN. 

In Founsure implementation, we have distinct object definitions for encoding, decoding, and repair operations. These objects have the trailer "*Obj" in common and include the same set of parameters in their object fields. For instance, both encoding and/or decoding functions accept \textit{EncoderObj} and/or \textit{DecoderObj} constructs as inputs. Similarly, $b$ and $k$ variables can be accessed using the standard way \textit{EncoderObj.sizesb} and \textit{EncoderObj.sizek}.

Each encoding/decoding object is associated with a seed value (\textit{EncoderObj.seed}\footnote{The default value selected for the see is 1389488782 which is experimentally observed to give good recovery performance.}) from which other seed values and the local sets of data chunks are pseudo-randomly generated. Each coding chunk within \textit{EncoderObj} and \textit{DecoderObj} has their own unique ID. These IDs are used to identify the erased coding chunks. The seed value is used by the pseudorandom generator to create a sequence of integers. These integers form the basis of coding chunk degree number assignment and the selected data chunks for coding chunk computations. These numbers are stored as part of the object and can be regenerated using the same initial seed number followed by the regular recurrence relationship. Let us assume we have \textit{s} number of output files (failure domains), then we use the default value \textit{EncoderObj.seed} + $i$ as the seed of the $i$th output file with $0 \leq i < \textit{s}$. 

\begin{figure}[t!]
\centering
\includegraphics[width=0.8\textwidth]{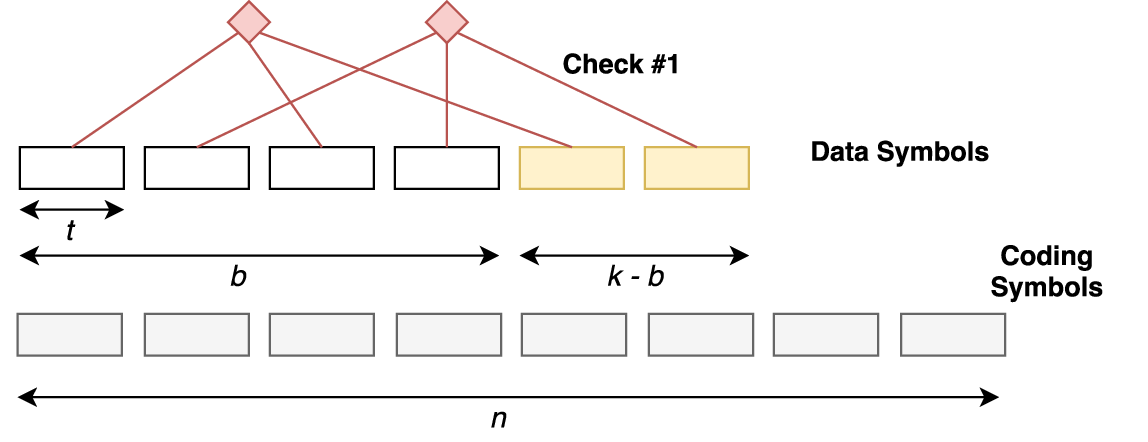}
\caption{Founsure symbols are $t$ bytes each and a precode is applied to find check \#1 equations. We add $k-b$ extra symbols to satisfy check equations as shown. Founsure encoding engine generates $n$ coding chunks from $k$ chunks as shown.}
\end{figure}

Each check node $c$ is associated with a degree number $c_d$ (chosen according to an appropriate degree distribution $\Omega(x) = \sum_i \Omega_i$ where $\Omega_i$ is the probability of choosing degree $i$) and $c_d$ data node neighbors are selected to be involved in final symbol computation. The degree distribution $\Omega(x)$ is typically selected to minimize the coding overhead. For instance, the following degree distribution is proposed for Raptor codes \cite{amin}
\begin{eqnarray}
&& \Omega(x) = 0.007969x + 0.49357x^2 +0.16622x^3 + 0.072646x^4 + 0.082558x^5 \nonumber \\
&& \ \ \ \ \ \ + 0.056058x^8 + 0.037229x^9 + 0.05559x^{19} + 0.025023x^{64} + 0.003135x^{65} \nonumber
\end{eqnarray}
where $\Omega_1 = .007969$, $\Omega_2 = .49357$, $\Omega_3 = 0.16622$, $\Omega_4 = .072646$, $\Omega_5 = .082558$, $\Omega_8 = .056058$, $\Omega_9 = .037229$, $\Omega_{19}= .05559$, $\Omega_{64}= .025023$, $\Omega_{65}= .003135$.

However, Founsure does not necessarily minimize overhead. It may optimize overhead, repair bandwidth, and complexity at the same time. We recommend choosing degree distributions that will give us a good trade-off point between these three objectives. A systematic optimization procedure to achieve the desired operating point is the subject of further investigation. Although there is no optimal point for all applications, Founsure is designed to be highly configurable to fit in different requirements and sensitivities of modern storage ecosystems.  

\begin{algorithm}[t!] 
\caption{Belief Propagation (BP) Algorithm (\textit{Skeleton})}\label{alg:bp}
\begin{algorithmic}[1]
\Procedure{BP}{$\textbf{B}$, $\mathcal{E}$, $maxit$}\Comment{Inputs: $\textbf{B} \in \mathbb{F}_2^{k \times n}, \mathcal{E} \subset \mathcal{N}$: The set of erasures.}
\State $\mathcal{N} = \{1,\dots,n\}$ \Comment{Initialize indexes.}
\State $\mathcal{K} = \{1,\dots,k\}$ \Comment{Initialize indexes.}
\State $\textbf{C} \gets \textbf{b}_{:,\mathcal{N} \backslash \mathcal{E}}$ \Comment{Find survival matrix.}
\State $\mathcal{F} \gets \mathcal{K}$ \Comment{Initialization: $\mathcal{F}$ holds the unrecovered indexes.}
\While {$\mathcal{F} \not= \varnothing \ \textbf{and} \  i < maxit$}
\For{$g=1,\dots,|\mathcal{N} \backslash \mathcal{E}|$}
\State $\textbf{c}_{\mathcal{K} \backslash \mathcal{F},:} \gets \textbf{0}$ \Comment{Zero-out rows for decoded symbols.}
\If{$weight(\textbf{c}_{:,g}) = 1$}  \Comment{Find degree-1 coding symbols.}
\State $\mathcal{F} \gets \mathcal{F} - \{f: \textbf{c}_{f,g} = 1 \textrm{ for } f \in \mathcal{F}  \}$
\EndIf
\EndFor
\EndWhile
\State \textbf{return} $\mathcal{F}$ \Comment{Return unrecovered indexes.} 
\EndProcedure
\end{algorithmic}
\end{algorithm}

We run {founsureDec} when we want to collect a subset of output data files and recover the input data file. Decoder is based on belief propagation algorithm  a summary of which is provided in {Algorithm \ref{alg:bp}}. BP function admits \textit{DecoderObj},  indexes of erasures $\mathcal{E} \subset \mathcal{N}=\{1,2,\dots,n\}$, the generator matrix of the base code $\textbf{B} \in \mathbb{F}_2^{k \times n}$ and a maximum number of iterations \textit{maxit}. In {Algorithm \ref{alg:bp}},  we use $\textbf{b}_{i,:}$ to refer to the $i$th row of $\textbf{B}$ and $\textbf{b}_{:,i}$ to refer to the $i$th column of $\textbf{B}$. Additionally, $\textbf{b}_{\mathcal{A},\mathcal{B}}$ refers to a matrix whose rows and columns are given by the rows and columns of $\textbf{B}$ indexed by the sets $\mathcal{A}$ and $\mathcal{B}$. The decoder utilizes the information contained in metadata file to generate (prepare) the contents of \textit{DecoderObj}, particularly the underlying coding graph. It works in a similar fashion to {founsureEnc} i.e., it reads the striped coding chunks, loads the buffer, and runs BP algorithm at most twice (once for the outer graph code and if need be, additional one for the inner Array LDPC precode) and recovers the $bt$ bytes at each turn. Finally, these bytes are written to decoded/recovered data file by calling standard kernel I/O commands. 

\begin{figure}[t!]
\centering
\includegraphics[width=92mm, height=95mm]{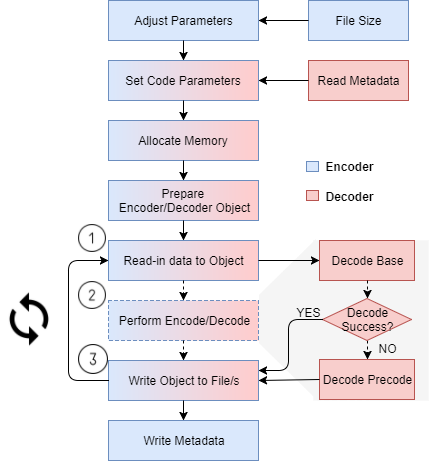}
\caption{Summary of software blocks of Founsure Encoder and Decoder.} \label{founsuresoftware}
\end{figure}

To summarize the software architecture of Founsure's encoding and decoding schemes, we provide Fig. \ref{founsuresoftware} to illustrate the order of software building blocks that take place to encode/decode the user data. The repair operation's architecture resembles a lot to this figure and hence omitted to save space. As can be seen, we used different colors for blocks to differentiate the encoding and decoding processes. Some of these blocks are used only by one of the functions. On the other hand, some blocks are used by both the encoder and decoder processes. 

Having all computation based on pseudorandomly selected chunks and carrying out these computations solely in terms of simple XOR logic has the cost of making the code non-optimal in terms of overhead (though it might be near-optimal through choosing appropriate degree distributions). If $n$ coding symbols are distributed over $s$ drives and when one of the drives fail, a subset of coding symbols are lost. To find what fraction of $f$-failure combinations can be tolerated for a given degree distribution, we provide a utility function {simDisk} that exhausts all possible combinations of failures to report which failures cases are tolerable by the code and which are not. Such a utility function is extremely useful for determining the reliability of the data protected by the Founsure library. 

\subsection{Check Equations and The Data Repair Process}
\label{}

We have three types of check nodes as mentioned before. We provide the details of such a checking process in this subsection.  
 
{\textbf{Checks \#1}}: These check equations are defined by the precode of the Founsure (for version 1.0, we selected an Array LPDC code family \cite{arrayldpc} for  efficient processing as given in {Algorithm \ref{alg:check1}}). Based on the selection of good precodes, the mathematical and coding parameter selections etc., the graph connections are automatically determined. Founsure includes a precode support based on a binary array LDPC code. Future releases of the library shall include external precode support which can be provided by the user using a preformatted input file. Please see the precoding subsection to find more information about the construction of these check equations. 

{\textbf{Checks \#2}}: These check equations are generated as given in  {Algorithm \ref{alg:cga}}. One of the special features of these checks is that only one neighbor is selected from the data nodes and the rest of the neighbors of the check node are from the coding nodes. This special feature can be used to partially decode the input data without running the complete decoder and reconstruct the unnecessary parts of the input data. An application of this could be securely stored multimedia source in which the Region of Interest (RoI) can be directly reconstructed using this type of check equations.

\begin{figure}[b!]
\centering
\includegraphics[width=0.6\textwidth]{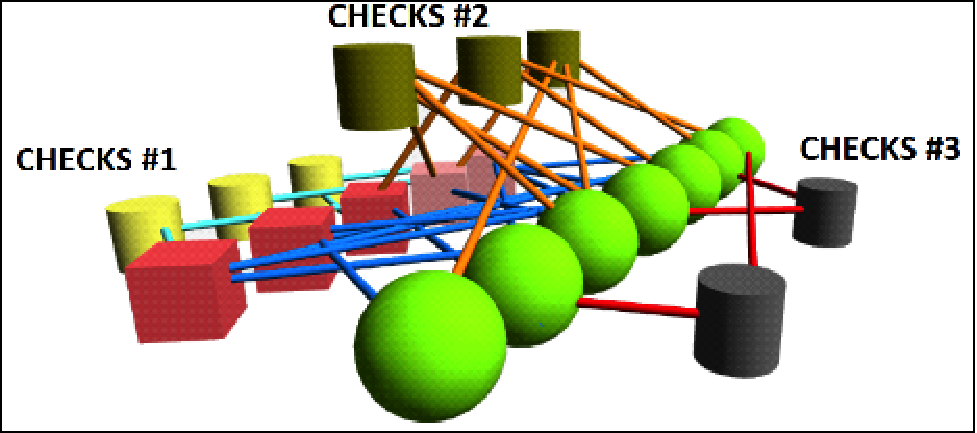}
\caption{Founsure graph code is three dimensional bipartite graph with three sets of check equations.}
\end{figure}

{\textbf{Checks \#3}}: These check equations are generated as given in {Algorithm \ref{alg:cga}}. These checks form the local groups based on the coding nodes. These checks are primarily used to repair the permanently erased, long-time unavailable, or unresponsive coding nodes in case of hardware, software, and network failures. 

\subsection{Precoding Process - Generation of Check \#1 equations}

A $(b,k,n)$ Founsure code takes $b$ data symbols (a total of $bt$ bytes) and initially generates $k-b$ check \#1 parity symbols based on the binary array LDPC encoding \cite{arrayldpc}. This special choice of array LDPC codes enables efficient encoding operation (linear with blocklength) and improves the complexity performance of the overall Founsure library.

\begin{algorithm}[t!]
\caption{Array LDPC Checks (check \# 1)}\label{alg:check1}
\begin{algorithmic}[1]
\Procedure{checkone}{$b$, $k$}\Comment{Inputs: $b,k$}
\State $\textbf{L}^{(1)} \gets \textbf{0}_{k \times k-b}$ \Comment{Initial zero matrix}
\State $p\gets$ largest\_prime\_factor($k$)
\State $k^\prime \gets k/p$
\State $j^\prime \gets  k^\prime - b/p$
\For{$j=0; j<j^\prime; j++$}
\For{$i=0; i<p; i++$}
\For{$m=1; m<k^\prime-j^\prime+1; m++$}
\State $\textbf{l}^{(1)}_{i+jp,m-1} \gets k^\prime p - (j^\prime - j + m - 1)p - (m(j^\prime-j-1) - i - 1 + p) \pmod p - 1$
\EndFor
\EndFor
\EndFor
\State \textbf{return} $\textbf{L}^{(1)}$ \Comment{ $\textbf{L}^{(1)}$: check \#1 sets.}
\EndProcedure
\end{algorithmic}
\end{algorithm}

The procedure outlined in {Algorithm \ref{alg:check1}} uses a generic function largest\_prime\_factor(.) which chooses the largest prime factor of the argument. The rate of the array LDPC is defined as $r_{LDPC} = b/k$. The user can choose any $k$, $n$ and $r_{LDPC}$ and hence we can calculate the appropriate $b = \lfloor kr_{LDPC} \rfloor$. Let $p =$ largest\_prime\_factor($k$), we may not be able to get the quantity $\lfloor kr_{LDPC} \rfloor / p$ equal to an integer. We can use the floor function to get an estimate of $j^\prime$. However, the array LDPC code performance is heavily dependent on $k^\prime$ and $j^\prime$ values and there is no array LDPC code for all $(k,r_{LDPC})$ pairs. If ($j^\prime,k^\prime$) pair are small, the code performance is observed to be pretty bad. For this reason, we provide an algorithm that reasonably chooses a good performing array LDPC code and satisfies (within some error margin) the user-provided parameters $k,n$ and $r_{LDPC}$ at the same time. One can see the chosen parameters by adding "-v" flag at the end of Founsure main functions. 

Let us define the following system parameters. After that, we shall formally provide the algorithm that determines the closest good-performing array LDPC code for the user-provided parameters $k,n$ and $r_{LDPC}$. These system parameters with their default values are defined in "parameter.h" file and can easily be modified. 

\begin{itemize}
\item DIFF\_TH: Allowed error threshold between the estimated and user provided $b$ values.
\item RRATE\_TH: Allowed error threshold between the estimated and user provided precode rate.
\item RED\_BYTE\_TH: Allowed redundant zero bytes to be appended at the end of the file for parameter consistency.
\item RAND\_WIN\_MAX: Random number search window maximum value.
\item RAND\_WIN\_MIN: Random number search window minimum value. 
\item ARRAY\_MIN\_JJ: Minimum Array LDPC "\textit{$j^\prime$}" parameter value.
\item ARRAY\_MIN\_KK: Minimum Array LDPC "\textit{$k^\prime$}" parameter value.
\item TRIES\_TH: Threshold on the number of tries before incrementing DIFF\_TH and RRATE\_TH. 
\item DELTA\_DIFF\_TH: Step size increment for DIFF\_TH
\item DELTA\_RRATE\_TH: Step size increment for RRATE\_TH. 
\end{itemize}

Next, we provide the algorithm that returns the estimated values of $b$, $k$, and the redundant number of zeros ($redundantzeros$) that need to be appended to the input user data.  The algorithm admits four inputs, namely $r_{LDPC}$, $filesize$, $b$ and $t$. Initial values of system parameters shall be set by "parameter.h" file and are changed locally within the function implementing {Algorithm \ref{alg:app}}. 

\begin{algorithm}[t!]
\caption{Adjust Parameters with Precode (APP)}\label{alg:app}
\begin{algorithmic}[1]
\Procedure{app}{$r_{LDPC}$, \textit{filesize}, $b$, $t$}
\State $\hat{b} \gets b, b \gets 0, k \gets 1, k^\prime \gets 1e9, j^\prime \gets 1e9, p \gets 2, blocks \gets 1, tries \gets 0, iter \gets 0$
\While {(|$b-\hat{b}$| > DIFF\_TH || |b/k - $r_{LDPC}$| > RRATE\_TH || $k^\prime > p$ || $j^\prime > p$ \\ || $k^\prime < j^\prime$ || |$blocks \times t \times b $ $-$  \textit{filesize}| > RED\_BYTE\_TH}
\State $k \gets \lfloor \hat{b}/r_{LDPC} \rfloor + rand() \mod$(RAND\_WIN\_MAX + RAND\_WIN\_MIN) - RAND\_WIN\_MIN
\State $p \gets$ largest\_prime\_factor($k$)
\State $k^\prime \gets k / p, j^\prime \gets k^\prime - \hat{b}/p$
\If{$j^\prime <$ ARRAY\_MIN\_JJ}
\State $j^\prime \gets 2$
\EndIf
\State $b \gets (k^\prime - j^\prime)p$
\If{$b > 0$}
\State $blocks \gets$ \textit{filesize} / $tb$ + 1
\State $tries++$
\If{$tries >$ TRIES\_TH}
\State DIFF\_TH $\gets$ DIFF\_TH + DELTA\_DIFF\_TH
\State RRATE\_TH $\gets$ RRATE\_TH + DELTA\_RRATE\_TH
\State $tries \gets 0$
\EndIf
\EndIf
\State $iter++$
\If{$iter > 1e7$}
\State print error and exit.
\EndIf
\EndWhile
\State $blocks \gets$ \textit{filesize} / $(tb)$ + 1
\State $redundantzeros \gets bt \times blocks$ - \textit{filesize}
\State \textbf{return} $b,k,redundantzeros$
\EndProcedure
\end{algorithmic}
\end{algorithm}

\subsection{Generating Information for Efficient Repair}


To efficiently repair the lost data, we need to extract repair information from the underlying graphical content of the generated Founsure code. We observe that check \#3 nodes are the most suitable node type for the repair process since it establishes a direct relationship between the coded symbols. It is not hard to see having more of these node types (created independently or dependently) gives alternative ways of repairing a given node in case of different combinations of node failures happening in the communication network. In other words, the more of these check types we find, the more potential we have for the regeneration of the lost coded chunks. With regard to this observation, we can use two techniques to increase the number of check \#3 type information based on check \#1 and check \#2 equations: 
\begin{itemize}
\item The first method is as follows. We identify the coded nodes with degree one (say we have $M$ of those nodes). Identify their data node neighbors. There are $M$ check \#2  equations that connect these data nodes with the coding nodes. Since the corresponding $M$ coding nodes carry the same information, we can use these check \#2 type check equations as additional check \#3 check equations. Note that since check \#2 and check \#3 equations are derived from the same base graph, this technique is likely to generate already existent local recovery groups or local groups that can be derived from existent local groups for the coding nodes.  
\item The second method is as follows. Note that check \#1 is user-defined although subject to a predefined structure. This defines local recovery groups over the data nodes. Since each data node is linked to local recovery groups of coded nodes through check \#2, we can use this relationship to derive check \#3 type check equations. For example, suppose we have the following check \#1 local recovery group defined for data nodes $D_0$, $D_1$ and $D_2$: $(D_0, D_1, D_2)$. Also, suppose that we have the following check \#2 equations:
\begin{eqnarray}
\Rightarrow D_0 &=& C_0 \oplus C_1 \oplus C_2 \\
\Rightarrow D_1 &=& C_1 \oplus C_{12} \oplus C_{17} \oplus C_{20} \oplus C_{99} \\
\Rightarrow D_2 &=& C_0 \oplus C_{21} \oplus C_{99}
\end{eqnarray}

Thus, we can find a check \#3 type equation given by $(C_{2}, C_{12}, C_{17}, C_{20}, C_{21})$ by observing the following equivalence,
\begin{eqnarray}
C_{2} \oplus C_{12} \oplus C_{17} \oplus C_{20} \oplus C_{21} = D_0 \oplus D_1 \oplus D_2 \label{EqnsetXOR}
\end{eqnarray}

Note that since this technique uses check \#1 equations, it is likely to generate distinct check \#3 local recovery groups and help improve repair performance dramatically.  These additional check \#3 local recovery groups (for instance the operation of Equation (\ref{EqnsetXOR})) are efficiently computed by the set union function \textit{setXOR} given in "encoder.c" file. 
\end{itemize}

\subsection{Algorithm for Jointly Generating Check \#2 and Check \#3 Equations}

We propose a heuristic algorithm to generate check \#2 and check \#3 equations at the same time for efficiency.  This algorithm uses XOR operation ($\oplus$) to sparsify  the generator matrix $\textbf{B}$. If $\textbf{B}$ is full rank, i.e., $rank($\textbf{B}$) = k$ then the algorithm is guaranteed to converge successfully. This is due to elementary matrix row operations shall generate $n-k$ zero columns for a full rank $\textbf{B}$ and hence the algorithm will leave the main while loop and generating a modified $\textbf{B}$ with column weights $k$ ones and $n-k$ zeros. In mathematical terms at the end, we should be able to find a permutation matrix $\textbf{P}$ such that $\textbf{BP} = [\textbf{I}_{k \times k} \ | \ \textbf{0}_{k \times n-k}]$. Also, $\tilde{\textbf{L}}^{(2,3)} = \textbf{L}^{(2,3)}\textbf{P}$ shall hold all the local recovery sets i.e., check \#2 equations in the first $k$ columns and check \#3 equations in the last $n-k$ columns. We can express different types of checks as the union of all check \#2 and check \#3 equations as given by
\begin{eqnarray}
\bigcup_{i < k} (D_i, \{C_s : {\tilde{\textbf{l}}^{(2,3)}_{s,i} \not = 0}\}) \ \ \cup \ \ \bigcup_{i \geq k} (\{C_s : {\tilde{\textbf{l}}^{(2,3)}_{s,i} \not = 0}\})
\end{eqnarray}
where $\tilde{\textbf{l}}^{(2,3)}_{s,i}$ denote $s$th row and $i$th column entry of $\tilde{\textbf{L}}^{(2,3)}$. In {Algorithm \ref{alg:cga}} we provide the details of the algorithm using pseudocode. We use a simple function zero\_columns(.) that finds the number of nonzero columns of the matrix in the argument. Since $\textbf{B}$ is typically sparse, we use a sparse representation of matrices in the library implementation for efficient memory utilization. We also order the local recovery groups based on their cardinality i.e., the set with the smallest cardinality comes first. Such an arrangement helps us reduce the repair/update complexity since the repair function processes local groups in sequential order. 

\begin{algorithm}[t!]
\caption{Check Generation Algorithm (CGA)}\label{alg:cga}
\begin{algorithmic}[1]
\Procedure{CGA}{$\textbf{B}$}\Comment{Input matrix $\textbf{B} \in \mathbb{F}_2^{k \times n}$}
\State $\textbf{L}^{(2,3)} \gets \textbf{I}_{n \times n}$ \Comment{Identity matrix}
\While{zero\_columns($\textbf{B}$) < $n - k$}\Comment{Number of zero-columns is checked.}
\State $temp \gets 0$
\For{j=0; j<n; j++}
\For{i=0; i<n; i++}
\If {$j \not= i$ and $2w(\textbf{b}_{:,j}) > w(\textbf{b}_{:,i})$}
\If {$w(\textbf{b}_{:,j} \oplus \textbf{b}_{:,i}) < w(\textbf{b}_{:,j})$}
\State $\textbf{b}_{:,j} \gets \textbf{b}_{:,j} \oplus \textbf{b}_{:,i}$
\State $\textbf{l}_{:,j}^{(2,3)} \gets \textbf{l}_{:,j}^{(2,3)} \oplus \textbf{l}_{:,i}^{(2,3)}$
\State $temp \gets 1$
\EndIf
\EndIf
\EndFor
\EndFor
\If {temp = 0}
\State \textbf{break};
\EndIf
\EndWhile\label{euclidendwhile}
\State \textbf{return} $\textbf{B}, \textbf{L}^{(2,3)}$\Comment{\textbf{B}:Check type, $\textbf{L}^{(2,3)}$:Local sets.}
\EndProcedure
\end{algorithmic}
\end{algorithm}

\subsection{Management of Check \#2 and Check \#3 Equations and the Generation of \textit{<filename>\_check.data} File for Efficient Repair/Update Process}

Check \#1 equations are determined through a binary array LDPC code as explained before. The user-defined number of equations are selected from a set of precode rates based on the reliability imposed by the application. The graph connections are deterministic and given by the constraints of the array code. 

Unlike check \#1, check \#2 and check \#3 are determined pseudo-randomly by the Founsure base code. Based on the generator matrix of the code \textbf{B}, {Algorithm \ref{alg:cga}} is run to determine $n$ equations. If the algorithm converges, then we should have $k$ equations for check \#2 type and $n-k$ equations for check \#3 type. The algorithm produces a correct set of local equations (sets) but does not guarantee those equations to be independent. In generating those equations, we do not employ any matrix inversions (which is quite costly for large size matrices) to find check equations and hence we trade off the efficiency by performance. The function that generates check \#2 and check \#3 local recovery equations is the utility function {genChecks}.

The function {genChecks} assumes that a metadata file is already generated by a previous run of the encoder {founsureEnc}. Hence {genChecks} generates check groups and modifies the \textit{meta\_data} file (appends the size of check data in terms of \textit{sizeof(int)} bytes at the end of the metadata file if “-m” flag parameter is True). The check information is stored in another binary file called \textit{<filename>\_check.data}. This file stores an integer array with a specific format. The reason for introducing a format is to use bulk read/write capabilities of \textit{fread} and \textit{fwrite} C library functions which will make kernel’s  I/O performance acceptable. 

The proposed format in this study is pretty straightforward and can be improved. We use flag bits to differentiate between the two distinct check equations. Thus, the integer value of the first \textit{sizeof(int)} bytes in  \textit{<filename>\_check.data} is either 0 or 1. 

\begin{itemize}
\item If it is 1 (Check \#2), then the next integer value (next \textit{sizeof(int)} bytes) gives the data symbol index which is involved within a local recovery group whose degree is given by the following integer (next \textit{sizeof(int)} bytes). This degree also indicates the next “degree” number, i.e., the number of integers to be read as part of one local recovery group for the coded symbols.
\item If it is 0 (Check \#3), then the next integer value (next \textit{sizeof(int)} bytes) gives the degree number i.e., the number of integers to be read as part of one local recovery group for coding symbols.
\end{itemize}

The nice thing about {Algorithm \ref{alg:cga}} is that if it converges, then all of the data symbols are covered exactly by one particular Check \#2 local recovery equation. Let us provide an example to illustrate the working principle and suppose that we have the following integer array stored in \textit{<filename>\_check.data}:
\begin{equation}
0 \ \ 4 \ \ 13 \ \ 56 \ \ 17 \ \ 66 \ \ 1 \ \ 19 \ \ 2 \ \ 11 \ \ 13 \ \ 0 \ \ 2 \ \ 39 \ \ 88 \ \dots  \nonumber
\end{equation}

If we decode this integer array, we will be able to say that the first local set is of type check \#3 and this set has four elements. In other words, 13th, 56th,17th, and 66th coding symbols form a local recovery group i.e., their binary sum should produce all-zero content. The next local set belongs to check \#2 and the associated data symbol index is 19. This data symbol along with 11th and 13th coded symbols (two coding symbols) forms a local recovery group. This way we can decode the whole integer array stored in <filename>\_check.data. If the algorithm converges, there should be $k$ leading 1’s and $n-k$ leading 0’s in the integer array not necessarily written in sequential order. Note that the total number of integers contained in the array is given by 
\begin{eqnarray}
N = \sum_{c=0}^{n-1}L_c + k + 2n \label{memorysize}
\end{eqnarray}
where $L_c$ is the total number of elements in check \#2 (excluding the data symbols) and check \#3 indexed by $c$. Note that even if the algorithm does not converge, the maximum memory occupancy possible is $N \times$ \textit{sizeof(int)} bytes. So it is sufficient to allocate the size of memory given by Equation (\ref{memorysize}) for the file without encountering a segmentation fault. 

\begin{figure}[t!]
\centering
\includegraphics[width=0.8\textwidth]{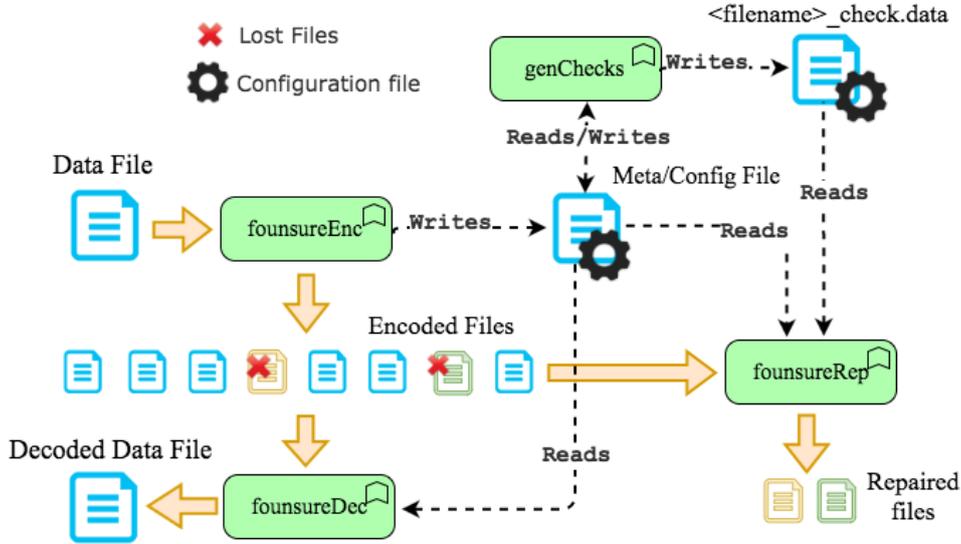}
\caption{Summary of interactions between different functions of Founsure.} \label{founsurefunctions}
\end{figure}

Fig. \ref{founsurefunctions} summarizes how different functions of Founsure interact with each other, what other metadata is generated/used, and what type of read/write permissions are granted to each of these functions for the proper operation of the Founsure library. 

\subsection{Reading/Formating the contents of \textit{<filename>\_check.data} file}

When the repair process is initiated, memory allocation, and repair object (\textit{RepairObj}) preparation, begins. The main repair engine shall look for \textit{<filename>\_check.data} under $/Coding$ directory. If it finds one and if the metadata is appropriately formatted (after a successful format check), it will read-in the metadata and format the check \# 2 and check \# 3 equations for the preparation of \textit{RepairObj}. A bulk read kernel call is performed and all the content is transferred to memory (inside the buffer pointed by $content2read$). Since Founsure's decoding, repair, and update operations are solely based on the BP algorithm, it sequentially searches only one unknown over the available local sets in a loop. To reduce the computation and bandwidth, the repair/decode process must use small size check \# 3 equations first so that we do not have to run through the end of the loop to complete the overall repair process. Founsure implementation extracts check \# 3 equations from  the buffer ($content2read$) using the standard $qsort(.)$ function and then fills in the  appropriate fields of \textit{RepairObj}. The ordering can be enabled or disabled for check  \# 2 and \# 3 equations using parameters ORDER\_CHECK\_2 and ORDER\_CHECK\_3 in "parameter.h" file. 

\subsection{Update Process}

An update process is about making the existing Founsure code stronger or weaker by either generating more redundancy (in case of increased failures or wear-out) or taking away unwanted redundancy (in case of using more reliable devices for storing information). If we would like to make the existing code weaker, it would not be hard. We just need to modify the metadata accordingly and erase the redundancy manually. Founsure does not automatically erase files and leave them to upper layer software management. So for the rest of this section, updating the code structure would mean making the code stronger. 

The desirable features of a generic update process can be listed as follows.
\begin{itemize}
    \item An update process should minimize the modification of the data generated by the encoding operation. 
    \item An update process should generate extra redundancy consistent with the encoded data with minimum processing effort.
    \item An update process should have a minimum limit on the extent of extra redundancy that can be generated. 
    \item An update process should not violate the rules set by the encoding and decoding processes. 
\end{itemize}

Founsure's update mechanism poses no modification changes to the already encoded data. In that sense, its update process functions as ideal as possible.  Founsure update process is tightly related to the repair process. This is mainly because updating a code is about repairing the missing blocks of information to help increase the reliability of data. We call {genChecks} to update the current code using the flag `-e'. There must be valid metadata associated with the code at the time {genChecks} is called. The code update process will rewrite $n$, the number of bytes used for the integer array due to check \#2 and check \#3 equations and update \textit{<filename>\_check.data}. Hence the repair process uses the metadata (the rule set) generated by the previous runs of the encoder/decoder pair. This series of modifications do not make any changes to the existent data/coding chunks. To trigger/sync changes with data, we finally need to call {founsureRep} function with the appropriate file name. Since the existent data is only read and we use minimum cardinality, local recovery groups, while updating, the processing effort is minimized. We finally note that since Founsure is based on fountain-like codes, there is no practical limit to the number of coding symbols that can be generated. As can be seen, the update functionality of Founsure is designed and implemented in the observation of desirable features listed above.



\section{Advanced Features}
\label{}

Although many advanced features of Founsure are described in previous sections, we have two more important implementation-specific advanced features that make Founsure's performance stand out. 

\subsection{Shared Memory Parallelism}

In shared-memory multiprocessor architectures,
threads can be used to implement parallelism. The shared-memory standard openMP is a high-level and portable interface that makes it
easier to use multi-threading capability and obtain satisfactory
performance improvements. Many erasure coding libraries such as Jerasure 2.0 \cite{Plank} has encoding/decoding engines that comprise independent “for” loop iterations and
hence possess huge potential for multi-threaded processing. Multi-threaded implementations
of Jerasure 2.0 are studied in \cite{suayb_siu} and \cite{suayb_blacksea}. 

As can be seen in Fig. \ref{founsuresoftware} for {founsureEnc}, the software consists of three stages executed in a loop. However, two of these stages, namely reading the data into the \textit{EncoderObj} and \textit{DecoderObj} that stay on memory and writing the object contents to the persistent storage devices require kernel I/O calls. As a result of these calls, the performance will be inhibited by the throughput performance of the underlying storage devices I/O bandwidth. Thus, our focus is essentially the second stage, namely the pure encoding and decoding in which the data traverses only between CPU caches and the main memory. 

Founsure utilizes the shared memory approach standard OpenMP library directives to help use multiple threads to handle the workload of encoding, decoding, repair, and update operations in parallel. However, to use OpenMP directives effectively, we needed to implement encoding/decoding operations differently. We use `-m' flag to set the number of threads in the main functions of Founsure. This parameter can independently be assigned but we provide recommendations for picking out the right number of threads for each main function because selecting the wrong number could result in degraded performance. For instance, we recommend it to be equal to the number of failure domains (disks for instance) for {founsureEnc} so that each data block is generated by a different thread. Considering each output file gets written to a different disk or storage node, we can maximize the overall throughput of the system if `-m' and `-s' flags are set to the same number provided that the underlying CPU architecture supports that many concurrent threads. 

In {founsureDec}, remember that we use {Algorithm \ref{alg:bp}} to resolve the user data in an iterative manner. Suppose that for a maximum of convergent $t_m$ iterations, we decode a set of data symbols (also known as the ripple size) $\mathcal{G}_i$ at the $i$th iteration for $i\in{0,1,\dots,t_m}$. However, we note that a decoded data symbol $s \in \mathcal{G}_i$ might be using another symbol $h \in \mathcal{G}_i$ while decoding, which results in intra-iteration decoding dependency. If we let $j$th iteration to use only decoded symbols in $\cup_{i=0}^{j-1} \mathcal{G}_i$, this would lead to another decoded set sequence $\mathcal{\overline{G}}_0,
\mathcal{\overline{G}}_1, \dots, \mathcal{\overline{G}}_{\overline{t}_m}$ where $\mathcal{G}_0 = \mathcal{\overline{G}}_0$ and $\overline{t}_m > t_m$. Note that upon convergence, we should have $\cup_{i=0}^{t_m} |\mathcal{G}_j| = \cup_{i=0}^{\overline{t}_m} |\mathcal{\overline{G}}_j|$. Although this new delayed BP will converge late compared to original version with single thread, this observation is not necessarily true with multi-threads as $\mathcal{\overline{G}}_i$s can be computed by multiple threads because data symbols in $\mathcal{\overline{G}}_i$ are decoded completely independent of each other and decoding process for each only share data for read operations eliminating potential race conditions. 
We note that for a given $k$, if we increase the block length $n$ we would need less number of iterations i.e.,  smaller $t_m$  and $\overline{t}_m$ with larger ripple sizes in each iteration. Finally, we note that we have many calls of {Algorithm \ref{alg:bp}} for decoding independent partitions of the user data. Using a shared memory approach, we use multi-threading to compute $\mathcal{G}_i$ in parallel in each iteration. Thus, the larger is ripple size, the better would become the performance of our implementation. It is recommended to use more threads as the number of coding blocks $n$ increases. We also recommend testing the best number of threads for a given $n$ to find the optimal value because this number is heavily dependent on the degree distribution. Finally, the multi-threaded implementation of repair/update operations is similar to decoding since in both cases, we resolve the repaired/updated data using the BP algorithm.  For {founsureRep} function, the recommended number of threads equal to either the number of repaired coding blocks or the number of extra coding blocks generated out of an update operation. 

We have two functions that take advantage of the multi-core multi-threaded systems and carry out the main operations of Founsure in parallel. These functions are \textit{EncodeComputeFast\_mt} for performing the data encoding and generating output files simultaneously and \textit{runBP\_mt} which runs the BP algorithm as parallel as possible. In our revised BP implementation, we remove the intra-iteration decoding dependency (see also {Algorithm \ref{alg:dpg}}) by
\begin{itemize}
    \item allowing BP to proceed using only the decoded symbols in $\cup_{i=0}^{j-1} \mathcal{G}_i$ in $j$th iteration or decoding step,
    \item allowing only one particular (lowest-degree) coding symbol to decode each source symbol in $\mathcal{G}_j$,
\end{itemize}
where the latter eliminates the possibility of race conditions (double writes by different threads) and optimizes the complexity performance by reducing the number of XOR operations.
Note that as the number of failures increases, the number of coded symbols decreases, and hence finding the lowest degree coding symbol will not usually end up with much performance improvement. 
We finally note that, since different threads deal with different levels of workloads, we use dynamic scheduling of threads in openMP.

\subsection{Optimal Decoding Path Generation}

In this section, we assume that node degree and selection distributions of Founsure is determined by different requirements of the system. Also,  $\mathcal{DP}$ represents the set of source and coding symbol pairs in which the coding symbol is used to decode the paired up source symbol. Thus, in the case of convergence of BP, we should expect $|\mathcal{DP}| = 2k$. The elements of $\mathcal{DP}$ is found according to {Algorithm \ref{alg:dpg}}. A careful look at the algorithm reveals that the following line does the local optimization of finding the lowest-degree coding symbol that decodes a specific source symbol. 
\begin{eqnarray}
\mathcal{DP} \gets \mathcal{DP} \cup  \{(f,g):  \textrm{ For each } f \in \mathcal{F}, g \in  \overline{\mathcal{G}}_i \textrm{ s.t. $\textbf{c}_{f,g} = 1$ and $g_d$ is minimum. }
\end{eqnarray}

One another note about {Algorithm \ref{alg:dpg}} is that by keeping unrecovered symbols in $\mathcal{F}$, we do not allow the same source symbol to be decoded more than once. This leads to suboptimality but helps us with multi-threaded implementation since it will save us from dealing with race conditions that would otherwise be handled with time-consuming locks. Also comparing it with  {Algorithm \ref{alg:bp}}, we can observe that symbol decodings are done one iteration at a time and hence symbols that are decoded at a given iteration do not help with other symbols that could have potentially be decoded within the same iteration. This approach is adapted to help with the shared memory implementation of the previous subsection. 

\begin{algorithm}[t!] 
\caption{Decoding Path Generation (DPG) Algorithm}\label{alg:dpg}
\begin{algorithmic}[1]
\Procedure{DPG}{$\textbf{B}$, $\mathcal{E}$, $maxit$}\Comment{Inputs: $\textbf{B} \in \mathbb{F}_2^{k \times n}, \mathcal{E} \subset \mathcal{N}$: The set of erasures.}
\State $\mathcal{N} = \{1,\dots,n\}$ \Comment{Initialize indexes.}
\State $\mathcal{K} = \{1,\dots,k\}$ \Comment{Initialize indexes.}
\State $i \gets 0$, $\mathcal{DP} \gets \varnothing$
\State $\textbf{C} \gets \textbf{b}_{:,\mathcal{N} \backslash \mathcal{E}}$ \Comment{Find survival matrix.}
\State $\mathcal{F} \gets \mathcal{K}$ \Comment{Initialization: $\mathcal{F}$ holds the unrecovered indexes.}
\While {$\mathcal{F} \not= \varnothing \ \textbf{and} \  i < maxit$}
\State $\textbf{c}_{\mathcal{K} \backslash \mathcal{F},:} \gets \textbf{0}$ \Comment{Zero-out rows for decoded symbols.}
\State $\overline{\mathcal{G}}_i \gets \{g: weight(\textbf{c}_{:,g}) = 1 \textrm{ for } g=1,\dots,|\mathcal{N} \backslash \mathcal{E}|   \}$ \Comment{Find degree-1 coding symbols.}
\State $\mathcal{DP} \gets \mathcal{DP} \cup  \{(f,g):  \textrm{ For each } f \in \mathcal{F}, g \in  \overline{\mathcal{G}}_i \textrm{ s.t. $\textbf{c}_{f,g} = 1$ and $g_d$ is minimum.}  \}$
\State $\mathcal{F} \gets \{f: weight(\textbf{c}_{f,\mathcal{G}_i}) = 0 \textrm{ for } f \in \mathcal{F}  \}$ \Comment{Update $\mathcal{F}$}
\State $i \gets i + 1$
\EndWhile
\State \textbf{return} $\mathcal{F}$ \Comment{Return unrecovered indexes.} 
\EndProcedure
\end{algorithmic}
\end{algorithm}

\section{Illustrative Examples}
\label{}

In this section, we provide a set of commands to use encoding, decoding, repair, and update features of Founsure. Note that Founsure comes with man pages or you can always use ``-h" flag command for immediate help when you call Founsure functions. Moreover, these examples are also included in the GitHub page along with several performances and unit tests.

The following command will encode a test file \textit{testfile.txt} with $k=500$ data chunks with each chunk occupying $t=512$ bytes. The encoder generates $n=1000$ coding chunks using $d=$`FiniteDist' degree distribution and $p=$`ArrayLDPC' precoding. Finally, generated chunks are striped/written to $s=10$ distinct files for default disk/drive allocation under \textit{/Coding} directory\footnote{This directory naming is conventional and maintained for the legacy of Jerasure erasure code library.}. The flag ``-v" is used to output parameter information used during the encoding operation. Founsure encoder also generates a metadata file with critical coding parameters which will later be useful for decoding, repair, and update operations. Without appropriate metadata, Founsure cannot operate on files. \\ \\
\fbox{\parbox{0.98\textwidth}{
\texttt{\textbf{founsureEnc} -f  testfile.txt -k 500 -n 1000 -t 512 -d 'FiniteDist' -p 'ArrayLDPC' -s 10 -v}}} \\

Now, let us erase one of the coding chunks and run the Founsure decoder. The decoder shall generate a decoded file \textit{test\_file\_decoded.txt} under  \textit{/Coding} directory. You can use ``diff" command to compare this file with the original. \\ \\ 
\fbox{\parbox{0.98\textwidth}{
\texttt{\textbf{rm -rf} Coding/testfile\_disk0007.txt \\ \textbf{founsureDec} -f  testfile.txt -v}}} \\

One of the things we notice about {founsureDec} function is that it does not recover the lost drive data \textit{Coding/testfile\_disk0007.txt}, because this function is responsible only for the original data recovery process. In storage systems, however, we need to recover lost data to maintain acceptable data reliability.  In Founsure, it is extremely easy to initiate the repair (current version only supports exact repair at the moment) process by running the following command. \\ \\
\fbox{\parbox{0.98\textwidth}{
\texttt{\textbf{founsureRep} -f  testfile.txt -v}}} \\

This would trigger the conventional repair operation and first shall decode the entire data and then re-run partial encoding to generate the lost chunks. In addition, {founsureRep} outputs the pure computation speed as well as the bandwidth consumed due to repair.  We observe that conventional repair is a heavily time and bandwidth-consuming operation. In fact, due to non-optimal overhead, the number of bytes that need to be transferred for the conventional repair is a little larger than the size of the original user file. Founsure supports fast and efficient repair as well. In order to use this feature, one needs to modify the metadata file and create an extra helping data/file called \textit{testfile\_check.data} which shall contain information for fast repair. Details can be found later in the document. To make these changes, we primarily run {genChecks} function. 
Finally, we can re-run the repair function as before and you will realize from the comments pointed out that the function will be able to recognize that there is available information for fast/efficient repair and will run that process instead of switching to conventional repair. You should be able to observe the reduced bandwidth consumed by the repair operation. \\ \\
\fbox{\parbox{0.98\textwidth}{
\texttt{\textbf{genChecks} -f  testfile.txt -m 1 -v \\ \textbf{founsureRep} -f  testfile.txt -v}}} \\

We can also use {genChecks} to trigger `update' functionality. For example, let us assume that the system reliability is degraded due to drive wear and we want to generate an extra two drive-worth information, in addition to already generated 10 drive-worth information. We use `-e' flag to modify metadata file as well as \textit{testfile\_check.data} for the update operation. This shall change the code and all its related parameters. However, in order to apply it to encoded data, we shall use {founsureRep} to generate new coding chunks and output files. Alternatively, you can erase drive info as well by supplying negative values for `-e' flag. In this case, you do not need to call {founsureRep} because there is nothing to generate. You can simply erase corresponding drive chunks after you scale the system down. \\
\\
\fbox{\parbox{0.98\textwidth}{
\texttt{\textbf{genChecks} -f testfile -m 1 -v -e 2 \\ \textbf{founsureRep} -f  testfile.txt -v}}} \\


\section{Numerical Results and Impact}

To check the set-up, accuracy, functionality of the library, several tests are included with the software package. Besides, to measure the encoding/decoding speed and bandwidth consumed in case of a data repair, we included several performance tests as well. Our performance tests are run on a server system the details of which are given in Table \ref{cpu_properties}. To be able to draw a summary of the library performance, we provide Table \ref{performance_table} for quantification of Encoding/Decoding speed. In our test, we used a 64MiB file, and encoded data are spread across 10 disks equally. We have used multi-threading support and set \textit{-m} parameter set to 12. While decoding, we have removed 2, 3, and 4 disk worth of information before running the decoder. The \textit{-t} parameter is judiciously chosen in powers of two to enable hardware-friendly operation. No exhaustive optimization is carried out.  As can be seen, with almost a half code rate, we could achieve super-fast encoding and decoding speeds with the current implementation. 

\begin{table}[htp!]
  \centering
    \caption{Server System CPU features.}
    \label{cpu_properties}
  \begin{tabular}{|c|c|c|c|}
    \hline
    \multirow{2}{*}{Property} & \multicolumn{2}{c|}{Name:Intel Xeon CPU E5-2620} \\
    \cline{2-3}
    & Value & Explanation \\
    \hline
    Socket & 2 & \\
    Core & 12 & 6 in each socket\\
    Clock speed & 2.4 & GHz (Max.) \\
    Threads & 24 & 2 in each core\\
    Arc. & IA64 & X86\_64\\
    L1 cache & 32KiB & \\
    L2 cache & 256KiB & \\
    L3 cache & 15360KiB & \\
    Main memory  & $\approx$ 8GiB & \\
    \hline
  \end{tabular}
  \label{tablo}
\end{table}

\begin{table}[htp!]
\centering
\caption{Encoder/Decoder Performance in MB/sec.}
\label{performance_table}
\begin{tabular}{|l|l|l|l|c|c|c|}
\hline
\multicolumn{1}{|c|}{k} & \multicolumn{1}{c|}{n}  & \multicolumn{1}{c|}{rate} & \multicolumn{1}{c|}{t} & \begin{tabular}[c]{@{}c@{}}Encoder\\ Performance \\ (MB/sec)\end{tabular} & Failure \#                & \begin{tabular}[c]{@{}c@{}}Decoder\\ Performance \\ (MB/sec)\end{tabular} \\ \hline
                        &                         &                           &                        &                                                                           & 2 & 3291                                                                      \\ \cline{6-7} 
                        &                         &                           &                        &                                                                           & 3                         & 2852                                                                      \\ \cline{6-7} 
\multirow{-3}{*}{1036}  & \multirow{-3}{*}{2180}  & \multirow{-3}{*}{0.4752}  & \multirow{-3}{*}{1024} & \multirow{-3}{*}{2722}                                                    & 4 & 2340                                              \\ \hline
                        &                         &                           &                        &                                                                           & 2                         & 2668                                                                      \\ \cline{6-7} 
                        &                         &                           &                        &                                                                           & 3 & 2496                                              \\ \cline{6-7} 
\multirow{-3}{*}{10246} & \multirow{-3}{*}{21800} & \multirow{-3}{*}{0.47}    & \multirow{-3}{*}{512}  & \multirow{-3}{*}{1691}                                                    & 4                         & 2356                                                                      \\ \hline
\end{tabular}
\end{table}

While the execution performance of the library is quite attractive, we can also show that it is also bandwidth friendly when the data is repaired. We considered the case $k=10246$ and $n=21800$ with a 100MiB = 104,857,600 bytes file while all the rest of the parameters are the same as before. If we consider double disk failures, the conventional repair method requires us to transfer 108,267,520 bytes of data for the repair to be successful. This is a little over 100MiB as expected due to overhead sub-optimality of the code used in Founsure. On the other hand, if we use the improved repair scheme suggested in this study, we can achieve a maximum of 65,952,320 bytes of transfer for successful recovery, which is almost $2\times$ more efficient use of bandwidth over that of the conventional method. Note again that no optimization is performed in terms of degree distribution $\Omega(x)$ and advanced graph partitioning to minimize repair bandwidth. 

Finally in Table \ref{complibs}, we have run a simple test to compare the performance of Founsure against two of the most efficient and heavily used erasure coding libraries, namely Jerasure 2.0\footnote{https://github.com/ceph/jerasure} and ISA-L of Intel\footnote{https://github.com/intel/isa-l}, based on RS codes (using Cauchy and Vandermonde Matrix constructions). This time, we used a 1GiB = 1,073,741,824 bytes. We realize that the parameters $k$ and $n$ cannot be selected too large for these libraries, due to their algebraic constructions which makes it extremely complex to deal with such low-level sub-packetizations. We set $k=10$ and $n=20$ to simulate half code-rate RS codes which are closest to Founsure's $0.47$ code rate previously selected. Note that with this selection, the number of disks used to store the generated content can be 20 at most (Parameter $n$ also characterizes the number of disks). Having more disks to distribute data would make the presented encoding/decoding performances worse. On the other hand, the number of disks does not change the performance of Founsure. Parameters of Jerasure and ISA-L libraries are selected to give the best performance on the same system defined in Table \ref{cpu_properties}. The decoding of these libraries is set to decode $8$ blocks (4 disks worth information). We also know that there is no simple repair mechanism defined for these RS-based constructions in publicly available libraries and hence adopted the conventional decoding-based repair while we compare the bandwidth consumed for repairing a single failed disk content in Table \ref{complibs}. Finally, we also provided approximate best and worst-case time complexity of encoding/decoding for RS and fountain codes as available from other research works in literature \cite{complexityRS}.

\begin{table}[htp!]
  \centering 
  \caption{Performance comparisons between different erasure code libraries.}
\begin{tabular}{|l|l|l|l|l|}
\hline
Library Name &  \begin{tabular}[c]{@{}c@{}} Worst/Best Time \\ Complexity \end{tabular}  & \begin{tabular}[c]{@{}c@{}}Encoder\\ Performance \\ (MB/sec)\end{tabular}  & \begin{tabular}[c]{@{}c@{}}Decoder \\ Performance \\ (MB/sec)\end{tabular} &  \begin{tabular}[c]{@{}c@{}} Repair BW \\ (MB) \end{tabular}  \\ \hline
ISA-L        & $O(n^2)/O(n\log^2(n))$ \cite{complexityRS} & 1420           & 1390           & 1073,7         \\ \hline
Jerasure 2.0 & $O(n^2)/O(n\log^2(n))$ \cite{complexityRS} & 780            & 870            & 1073,7         \\ \hline
Founsure 1.0 & $O(n \log(n))/O(n)$  \cite{sarslan1} & 1620           & 2245           & 675,35         \\ \hline
\end{tabular}
\label{complibs}
\end{table}

To the best of my knowledge, Founsure is the most flexible erasure coding library that is open source and can be configured based on the requirements of the application. With the current software architecture, many more functionalities can be integrated such as partial user data construction and advanced error detection for failure localization. Also, as numerical results suggest even if many more optimizations are possible to make the performance better, the current version's performance in terms of execution speed and repair bandwidth still stand out. Founsure has highly parallel architecture and lends itself to parallel programming. Unlike Founsure, existing research is mostly focused on overhead optimal designs using the inherent nature of the parallel hardware. Originally, Founsure is developed for data storage systems, it can simply be adapted to packet-switched networks in which the underlying channel is erasure channel or sporadic erasure channels \cite{liva2010}. With advanced features such as error detection, erased content reconstruction, multi-threaded support, advanced decoding, Founsure can further be used for error correction which could open up more areas of applications such as image reconstruction and data protection over noisy communication channels. Consequently, we believe that Founsure could be a strong candidate to be used for any system that secures data protection and recovery in one way or another. 

\section{Conclusions}
\label{}

In this work, we have developed and presented an erasure coding library that can be used to operate on various points of the trade-off between computational complexity, coding overhead, and repair bandwidth. For example, through the right selection of coding parameters, the Founsure library can be used to save storage space and minimize the data storage overhead. On the other hand, by allowing some overhead again through tweaking parameters, Founsure can reduce the data repair bandwidth. Unlike previous software packages, such freedom of parameter selections makes Founsure library more application-centric and configurable for future generation reliable system design.

\section*{Acknowledgements}
\label{}

I would like to thank The Scientific and Technological Research Council of Turkey (TUBITAK) and Quantum Corporation which provided the support for maturing ideas, and necessary hardware platforms for extensive testing. 






\section*{Current code version}
\label{}

\begin{table}[!h]
\begin{tabular}{|l|p{6.5cm}|p{8cm}|}
\hline
\textbf{Nr.} & \textbf{Code metadata description} & \textbf{Please fill in this column} \\
\hline
C1 & Current code version & 1.0 \\
\hline
C2 & Permanent link to code/repository used for this code version & \url{https://github.com/suaybarslan/founsure} \\
\hline
C3 & Legal Code License   & LGPLv3 \\
\hline
C4 & Code versioning system used & git \\
\hline
C5 & Software code languages, tools, and services used & C, Python, OpenMP, etc. \\
\hline
C6 & Compilation requirements, operating environments \& dependencies & Linux Runtime Environment and a C compiler ($\geq 4.8.4$). Python 2.3 or above. \\
\hline
C7 & If available Link to developer documentation/manual &  \url{https://github.com/suaybarslan/founsure/blob/master/tests/Founsure_1_0_User_Manual.pdf} \\
\hline
C8 & Support email for questions & arslans@mef.edu.tr \\
\hline
\end{tabular}
\caption{Code metadata (mandatory)}
\label{} 
\end{table}




\end{document}